# Acoustofluidic phase microscopy in a tilted segmentation-free configuration

Julián Mejía Morales, Björn Hammarström, Gian Luca Lippi, Massimo Vassalli, and Peter Glynne-Jones

## ARTICLES YOU MAY BE INTERESTED IN

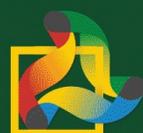

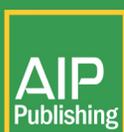

# Acoustofluidic phase microscopy in a tilted segmentation-free configuration

Julián Mejía Morales,[1,2,a)] 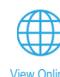 Björn Hammarström,[3,4] 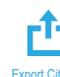 Gian Luca Lippi,[1] 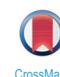 Massimo Vassalli,[5]
and Peter Glynne-Jones[3]

AFFILIATIONS

[1]Institut de Physique de Nice, Université Côte d'Azur, CNRS, 06560 Valbonne, France
[2]Department of Experimental Medicine, University of Genova, 16149 Genova, Italy
[3]Engineering Sciences, University of Southampton, SO17 1BJ Southampton, United Kingdom
[4]KTH Royal Institute of Technology, 100 44 Stockholm, Sweden
[5]James Watt School of Engineering, University of Glasgow, G12 8LT Glasgow, United Kingdom

[a)]Author to whom correspondence should be addressed: julian.mejia@inphyni.cnrs.fr

ABSTRACT

A low-cost device for registration-free quantitative phase microscopy (QPM) based on the transport of intensity equation of cells in continuous flow is presented. The method uses acoustic focusing to align cells into a single plane where all cells move at a constant speed. The acoustic focusing plane is tilted with respect to the microscope's focal plane in order to obtain cell images at multiple focal positions. As the cells are displaced at constant speed, phase maps can be generated without the need to segment and register individual objects. The proposed inclined geometry allows for the acquisition of a vertical stack without the need for any moving part, and it enables a cost-effective and robust implementation of QPM. The suitability of the solution for biological imaging is tested on blood samples, demonstrating the ability to recover the phase map of single red blood cells flowing through the microchip.

Published under license by AIP Publishing. https://doi.org/10.1063/5.0036585

## I. INTRODUCTION

Quantitative Phase Microscopy (QPM) is a computational approach that utilizes bright field optics to generate cell phase maps.[1,2] Each pixel of a phase map brings information on the optical thickness of the sample, a convolution between the actual thickness and the refractive index (RI). For an homogeneous material of known RI, the QPM technique provides a direct measure of the volume. QPM has demonstrated to be effective in many applications, and it offers significant advantages over traditional microscopy when imaging transparent phase objects, such as live cells in culture medium.[3]

Two main families of QPM implementation exist, either based on interferometry [as in Digital Holography (DH)[4]] or associated with the numerical processing of a stack of defocused bright field images [making use of the Transport of Intensity Equation (TIE)[2]]. The latter presents several advantages which render it simpler and less expensive than DH-based QPM:[4] less stringent mechanical stability and requirements on path length (since it avoids interference), computational simplicity, operation with partially coherent illumination sources, and the lack of constraints on hardware.

QPM is a powerful label-free technique offering high sensitivity, long term non-invasive imaging, and robust morphometric quantification of living cells.[2,5] For example, QPM has been effectively used to measure the RI and the osmotic-induced volume changes in erythrocytes,[3] providing a precise morphometric description of cells. The same approach on white blood cells allowed for the identification of robust biomarkers suitable for discriminating between healthy and diseased states.[6] Moreover, for nonhomogeneous materials, such as living cells, QPM can be seen as a contrast enhancement technique to highlight organelles and structures with different RIs. This perspective has been adopted to describe intracellular motility and dynamic evolution.[7,8] Furthermore, QPM has been used to detect cell parasites such as *Plasmodium falciparum* inside Red Blood Cells (RBCs).[9,10] Traditional bright field and phase contrast microscopy, including

Differential Interference Contrast (DIC) microscopy, is less suitable for identifying complex structures such as the mitochondrial networks inside the cell due to the low absorption and scattering contrast of the biological samples.[2] On the other hand, QPM enables such structure visualization, together with their dynamics, with high contrast.[11,12]

The main limit to the applicability of TIE-based QPM is the need for acquiring a stack of images across the focal plane with well defined focal offsets. Typically, this requires the mechanical translation of either the camera or the object,[13] introducing an expensive hardware component and requiring a longer and delicate imaging procedure. This is particularly detrimental for the implementation of high throughput imaging procedures for diagnostic purposes. Some solutions have been proposed to achieve high throughput TIE-based QPM, either implementing strategies for fast scanning of the focal plane using tunable lenses[14] or using smart optical configurations to obtain one-shot TIE imaging.[15–17] However, these approaches require rather expensive optical components or substantial modifications to the standard bright field microscopy apparatus.

In this work, we propose a simple and inexpensive technique for a fast, real-time implementation of TIE cytometry by combining a microfluidic channel, in a tilted configuration, with cell acoustic focusing. A transmission microscope is used to acquire images of the cells, levitated by the acoustic field, and translated by the fluid flow through the optical system's focal plane, where the latter forms a controlled angle with the acoustic field's focal plane located inside the microfluidic channel [cf. Fig. 1(b)]. Images are divided along the flow direction into stripes so that each stripe corresponds to a region of a similar amount of defocusing. Individual stripes are assembled into stacks for processing. Minimal registration is needed, as the imaging rate is matched to the flow rate such that objects appear in the same position in each stripe.

In a normal microfluidic channel, the parabolic flow profile would cause flowing cells to move with a wide distribution of velocities depending on their proximity to the walls. The acoustic focusing moves all cells to the same position in the flow profile, ensuring that they all move at the same speed and maintain a fixed arrangement within the focused plane.[18] The described configuration, therefore, allows for high throughput by retrieving the phase map of several cells at once, leading to a performance increase without special optical elements.

Cell orientation is also manipulated by the acoustic field. Previous work[19] has shown that in an acoustic planar standing wave, RBCs are caused to rotate such that the disk of the cell lies within the pressure nodal plane. In our setup, this creates a "flat" view of the cells which exposes a maximal area of each cell to the optics, and we conjecture that this leads to optimal information from the image compared with other possible orientations. Additionally, by aligning cells to the center of the flow channel, where the shear stress gradients are at a minimum and symmetrical, the cells are not induced to "tumble" as would be found in other arrangements.[20] This is important as the method described here relies on a lack of rotation between exposures.

Our configuration, thus, enables the realization of an inexpensive device usable for cytometry-based diagnosis, a highly desirable goal (particularly) in low-income countries, where access and quality of health care must be coupled to low costs.

## II. MATERIALS AND METHODS
### A. Chip production

The microfluidic chip used in this work is based on a previously proposed geometry.[18,21] In short, three layers of double sided adhesive transfer tape sheet (468MP, 3M, USA) were joined to produce a bond film $320 \mu m$ thick. Using a laser cutter, the joined bond film was cut into $5 \times 7.5$ cm pieces with a 6 mm wide channel running diagonally across. The cut pieces were sandwiched between two double width microscope slides ($1 \times 50 \times 75$ mm$^3$), and access holes to the microchannel of 1 mm diameter were drilled at both ends of the channel.

Underneath a portion of the fluid channel, a piezoelectric transducer of lead zirconate titanate (PZ-26, Ferroperm, Kvistgaard, Denmark) was attached with epoxy (Epotek-301, Epoxy Technology, Inc., USA). The size of the transducer was ($1 \times 25 \times 50$ mm$^3$), and a wrap-around electrode was created on the top-surface, in contact with the glass, using conductive silver paint (SCP Silver Conductive Paint, Electrolube Ltd., UK).

To drive the piezoelectric transducer, a TTi TG1006 signal generator was used in conjunction with a custom amplifier based around a high frequency op-amp (linear technologies, LT1210). Frequency was set at 2.286 MHz with a voltage amplitude of 2 Vpp to achieve the acoustic focusing. This results in an acoustic pressure node along the device's half-height center plane, which creates acoustic radiation forces that direct blood cells toward that plane.[18]

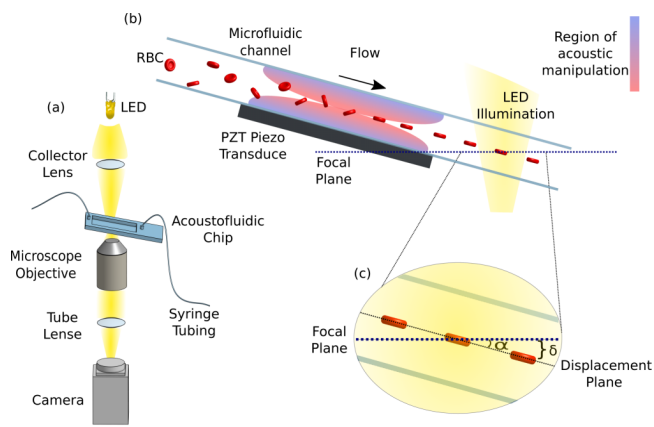

**FIG. 1.** Experimental configuration. (a) A pulsed light emitting diode (LED) illuminates the inclined microfluidic channel equipped with acoustic focusing. The images are collected by using a microscope and acquired with a digital camera. (b) A cross-sectional view of the microfluidic chip. The red blood cells (RBCs) initially travel in random orientations before entering the acoustic field created by the transducer. The radiation forces rotate the RBCs to be parallel to the chamber walls and align them into a plane in the device center where they move with uniform velocity. (c) The device tilt, relative to the optical axis, causes them to move diagonally through the focal plane defined by the microscope objective, thereby producing images with a controlled amount of defocusing.

### B. Microscopy setup

The microscopy setup in this work is based on an inverted transillumination microscope equipped with an Olympus LMPLFLN 50X Objective, $NA = 0.5$ and field of view (FoV), a square of side 0.53 mm [cf. Fig. 1(a)]. The microfluidic chip is positioned in the standard XY stage using a custom 3D printed frame. The tilt angle of the flow plane is controlled using glass slides as spacers mounted under one end of the chip; three 1 mm thick slides were used to achieve an angle of 3.5°, thus providing a total change in the field of view of $\Delta z \approx 15\,\mu m$.

With normal illumination, a particle traversing the field of view creates significant motion blur during the camera's exposure time. To avoid this effect, pulsed illumination was implemented to capture the cell images during short pulses of light. For this purpose, the original filament was replaced by a high intensity light emitting diode, LED (LZ1-00G102, LED Engin, CA, USA). This was driven by a custom driver circuit.[22] The duration of the pulses of light and their synchronization with the camera triggering were controlled by a microprocessor (Arduino Uno, Arduino). Pulse length was set to 10 $\mu$s.

### C. Blood samples

Rat's blood was mixed with lithium heparin at 20 IU/ml. The blood was centrifuged at 2000 rpm for 5 min and the plasma and buffy coat removed by using Pasteur pipet. 50 $\mu$l of the centrifuged blood was diluted with 5 ml of a hypotonic solution of PBS buffer (Sigma Aldrich #P4417) obtaining a cell count of $\approx 3 \times 10^3$ cells/$\mu$l, equivalent to dilute 3400 folds the mice blood's original samples. The diluted blood samples provided a suitable cell density for imaging, in our case 40–70 cells in each field of view (FOV). Experiments were conducted at room temperature (around 20 °C) and carried out within 48 h of blood collection.

A syringe pump (Harvard Apparatus Pump Elite 11) and a 10 ml glass syringe were used to pump the samples through the device at 200 $\mu$l/min. Our previous work[18] had shown that the pulsed illumination used here with 10 $\mu$s exposure time resulted in insignificant motion blur at the flow rates we demonstrate. The optical results presented here are, thus, not significantly dependent on flow rates, and with the RBCs near the center of the channel, we do not anticipate any significant shear stress induced changes in the RBCs.

### D. Image acquisition and processing

Images were acquired using a Hamamatsu Flash 4.0 C11440 monochrome camera set at 90 fps and 12 bits depth. All parameters were manually controlled, and any auto-balance feature was switched off. To eliminate artifacts due to partially uneven illumination and fixed objects in the field of view (dust and debris), an image of the static background $I_{BG}$ was extracted from the median over time of the first ten images. The adoption of the median (instead of the average) ensures that image features associated with particles passing through the frame are not given significant weight in the average (image averaging leads to particle's trails). The background image $I_{BG}$ was then subtracted from every acquired image $I_{CCD}$ and the difference $I = I_{CCD} - I_{BG}$ used to recover the phase.

Each image $I(t)$ was divided into five stripes $I_m(t)$, $m = -2, -1 \cdots 2$, of identical width $\xi = L/5$, where $L$ is the image's length along the direction of the flow [cf. Fig. 2(a)]. The floating objects are acoustically focused in the same plane so that they all move with the same speed $v$. If a series of five images is acquired with a delay $\tau = \xi/v$, the same objects imaged in a stripe $m$ at time $t$ will be in the stripe $m + 1$ at time $t + \tau$ but at a different vertical position. Thus, if we call $z_m(x, y)$, the distance from the focal plane of each point of the stripe $m$, we can write $z_m(x, y) - z_{m+1}(x, y) = \delta/5$, where $\delta$ is the maximum defocusing [see Figs. 1(c) and 2(a)]. Taking this into account, a virtual intensity stack $\hat{I}(x, y, z)$ can be constructed using the five intensity stripes $I_{m-2}(t - 2\tau) \cdots I_{m+2}(t + 2\tau)$ containing the same field of view at different heights. Little deviations might arise in the alignment of the stripes by simple translation, mainly due to optical aberrations. To correct for this deviation, the position of the stripes was finely tuned based on a strict registration procedure.[23]

### E. Solution of the TIE

The solution of the TIE, based on the two-dimensional Fast Fourier Transform (FFT), is fast and computationally simple.[4,24] First, the right hand side of Eq. (1), which represents the change of intensity along the $z$ axis, is computed by taking the forward derivative evaluated at the in-focus plane $z = 0$, the standard procedure in the QPM TIE-based approach.[12,25,26] Then, a two dimensional FFT is applied, to the obtained derivative, to retrieve the phase gradient $\nabla_\perp \phi(x, y)$. A periodic boundary condition was applied to the stack when calculating the 2D FFT, which was useful to convert the rectangular symmetry of the stack to a squared one. Indeed, a square symmetry is more favorable for 2D FFT calculations and is less prompt to artifacts of high frequencies. In addition, the 10th percentile of the frequencies with larger amplitude $|f(u, v)|$ was removed to eliminate the periodic noise, regardless of its high- or low-frequency nature. Finally, a second 2D FFT was applied to obtain the desired phase map $\phi(x, y)$.[14] The computational TIE solution was based on the Numpy Pyhton library.[27]

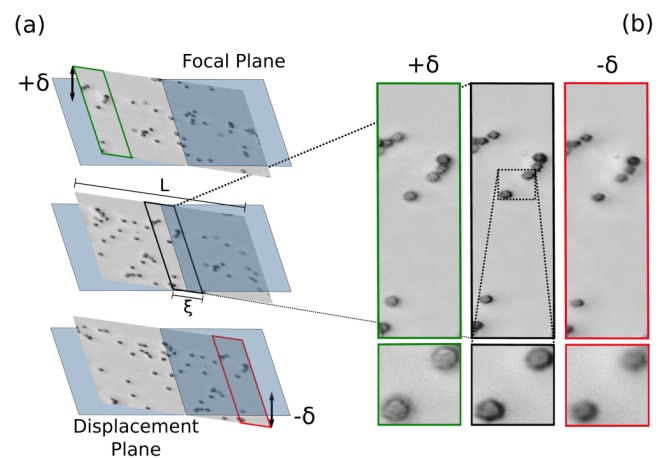

**FIG. 2.** Schematics of the experimental setup. (a) Acoustic focusing forces floating particles to flow along a plane (in gray) that is tilted with respect to the focal plane (in cyan). Samples in different regions of the image will be at different distances from the focal plane. If $z$ is the coordinate along the optical axis, objects in the central stripe (black frame) will correspond to $z = 0$, while objects in the first and last stripe (green and red, respectively) will sit at a higher distance from the focal plane, $z = \pm \delta$ for them. The image of length $L$ was divided in five stripes of identical width $\xi$. (b) Images from each strip, taken sequentially, to show the distribution of particles acquired at each vertical position, as defined by the stripes. The time series is converted into a series of vertical stacks by aligning the requisite stripes from images.

## III. RESULTS

The acquired $\hat{I}(x, y, z)$ stack can be exploited to calculate the optical thickness of the sample (phase reconstruction) using the TIE,[25]

$$\nabla_\perp [I_0(x, y) \nabla_\perp \phi(x, y)] = -k \frac{\partial I(x, y, z)}{\partial z}\bigg|_{z=0}, \quad (1)$$

where $k$ is the wavenumber, $\nabla_\perp$ is the in-plane gradient, $I_0(x, y) = \hat{I}(x, y, 0)$ is the image in the focal plane, $\phi(x, y)$ is the phase (to be reconstructed), and $\partial z$ indicates the derivative along the optical axis evaluated at $z = 0$. Under the assumption of a homogeneous and partially coherent illumination, this equation can be numerically solved to finally obtain a quasi-3D real-time reconstruction of the object. The numerical solution of the TIE has been described in Sec. II E.

The proposed approach has been adopted here to study a sample of murine blood. Figure 3(a) shows reference images of RBCs acquired with this method. For each sample, the in-focus bright field stripe (first column, $m = 0$) and the phase reconstruction (third column) are reported, together with the phase gradient (central column). The phase gradient $\nabla_\perp \phi(x, y)$ is calculated from the last column and offers a high contrast qualitative representation. The effectiveness of the method is highlighted in Fig. 3(b), where two phase profiles are shown, one obtained on a clump of cells (blue dotted profile) and the second on an isolated RBC (red dotted profile). This picture demonstrates the quantitative information carried by the TIE reconstruction. Aggregated cells (blue) are, in fact, higher than the single RBC (about twice as much) and present a rounded shape. In contrast, the red profile clearly highlights the expected donut-like shape of the isolated RBC, which is acoustically trapped in a planar configuration.

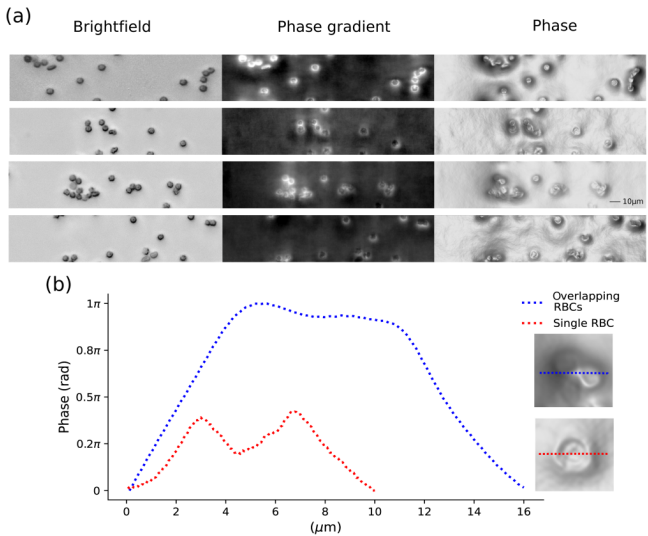

**FIG. 3.** (a) The left column shows the stripe of the brightfield image of the RBCs in the focal plane; the center column shows the gradient of the phase $\nabla \phi(x, y)$ and the right column shows the phase of the images $\phi(x, y)$. (b) Phase profile for a single RBC and for a clustered group of RBCs.

## IV. DISCUSSION AND CONCLUSION

A simple chip geometry was utilized, with a rectangular cross section where the width was much larger than the thickness. In this configuration, the flow was laminar with a parabolic velocity profile (objects close to the center of the chip flow faster than objects near the walls). This geometry was previously adopted to perform image cytometry on single algae, demonstrating a strong focusing efficiency, also resulting in a narrow velocity distribution of the floating objects.[28]

The adoption of a tilted configuration to obtain motion-free vertical stacks of floating objects has been previously suggested as a low-cost method to perform TIE-based QPM reconstruction.[25,29,30] Here, we propose a substantial improvement, provided by the integration of the acoustic focusing mechanism. Two main advantages are associated with this configuration. First, all objects flow through the field of view in the same plane so that an efficient stacking with controlled distances between planes is achieved for all objects, not only for those that were by chance in the correct region. The ability to finely control the distance between planes is crucial for the quality of the reconstruction, which is based on the calculation of the derivative. Moreover, focusing objects to a single plane means that they all move with the same velocity, and the motion can be efficiently approximated as a rigid translation. The majority of RBCs were observed to have rotated in the acoustic manipulation region to orient themselves within the pressure nodal plane and present a "flat" view.[19] Furthermore, due to the absence of significant shear stresses at the channel half-height, no significant cell rotation was observed between the successive exposures required to build up the QPM image. To reconstruct the virtual stack in this condition, we are not required to isolate and track the motion of individual cells, resulting in a more rapid and efficient image processing that can incorporate aggregates and objects of unexpected size.

The proposed device offers a low-cost approach to obtain physical maps associated with the refractive index and thickness of the sample that can be further manipulated and exploited for different aspects depending on the needs of the specific application. QPM maps can be rendered as 3D-like images of the sample, in which single objects can be effectively segmented using simple and fast threshold-based algorithms, thus enabling effective extraction of real-time morphological features (see image 3 and Ref. 31). Moreover, the same setup is suitable for addressing high throughput recognition of infected cells, where the presence of an intracellular parasite is known to directly impact the phase map, as in the case of malaria infections.[32] While QPM is primarily considered an imaging technique, it measures a physical quantity that relates to the refractive index, and this offers the possibility of using the system as a spectroscopic device for the identification of the sample's material properties, as recently proposed for the recognition of microplastics in sea water.[33] The method is based on the acquisition of image stripes that can accommodate more than one cell at the time (≈10 cells at the same time were mapped during experiments), thus increasing the overall potential throughput (in terms of cells/s).

In conclusion, the segmentation-free TIE-based setup proposed in this paper demonstrates a robust high throughput low-cost single cell in-chip QPM. The simple implementation offers a nonexpensive solution, suitable for environments where the cost is a core requirement, but where throughput and accuracy are mandatory. Examples include the identification of rare infected cells in tropical diseases,[34] screening applications in veterinary or food science,[35,36] or the identification of specific micrometric targets in environmental samples.[28,33]

## AUTHORS' CONTRIBUTIONS

J.M.M. and B.H. contributed equally to this manuscript.

## SUPPLEMENTARY MATERIAL

Supplemental material provides a detailed description for the reconstruction of virtual stacks for transport of intensity equation QPM implementation.

## ACKNOWLEDGMENTS

The authors would like to thank Dr. Stéphane Barland (UCA Nice) and Dr. Marco Sartore (ElbaTech SRL) for fruitful discussions. J.M.M. acknowledges the funding for international mobility France–Italy provided by the Université Franco Italienne (UFI, Project No. C2-1031) and the Mexican Council of Science and Technology (CONACyT) scholarship (No. 471712). P.G.J. gratefully acknowledges fellowship funding by the UK EPSRC (No. EP/L025035/1). This work has also been supported by the French government through the UCAJEDI Investments in the Future project managed by the National Research Agency (ANR) with Reference No. ANR-15-IDEX-01.

## DATA AVAILABILITY

The data that support the findings of this study are available from the corresponding author upon reasonable request.